\documentclass[preprint,aps]{revtex4}

\usepackage{epsf}
\usepackage{anysize}

\usepackage{amsmath}
\usepackage{amssymb}
\usepackage{amsfonts}
\usepackage{array}
\usepackage{dcolumn}
\usepackage{bm}
\usepackage{mathrsfs}
\usepackage{float}
\usepackage{graphicx}
\usepackage{hyperref}
\usepackage[latin1]{inputenc}
\usepackage{makecell,multirow,diagbox}

\newcounter{RomanNumber}

\begin{document}

\title{How to Synthesize Exceptional Points with Three Resonators}

\author{Re-Bing Wu,$^{1,3\dag}$, Yu Zheng,$^{2}$, Qi-Ming Chen,$^1$ and Yu-xi Liu$^{2,3}$ }
 \affiliation{$^1$Department of Automation, Tsinghua University, Beijing, 100084, China}
\affiliation{$^{2}$The Institute of Microelectronics, Tsinghua University, 100084, Beijing, China
 \affiliation{$^3$Center for Quantum Information Science and Technology£¬ BNRist), Beijing 100084,}
}

\date{\today}

\begin{abstract}
In non-Hermitian coulped-resonator networks, the eigenvectors of degenerate eigenmodes may become parallel due to the singularity at so-called Exceptional Points (EP). To exploit the parametric sensitivity at EPs, an important problem is, given an arbitrary set of coupled resonators, how to generate a desired EP by properly coupling them together. This paper provides the solution for the case of three resonators. We show that all physically admissible EPs can be realized with either weakly coupled linear networks or strongly coupled circular networks, and the latter type of EPs has not been reported in the literature. Each admissible EP eigenvalue can be realized by two and only two resonator networks, and the formulas for calculating the required coupling constants are provided. The characteristics of these EPs are illustrated by the change of transmission spectra near them, which verify the enhanced sensitivity induced by the singularity of EPs.
\end{abstract}

\maketitle

\section{Introduction}\label{sec:intruduction}

The concept of exceptional points (EP) was firstly coined by Kato \cite{Kato1966} in 1966. An EP$_n$ is referred to a linear system whose $n$ eigenvectors coalesce as well as the corresponding $n$ eigenvalues. Mathematically, EP$_n$ corresponds to an $n$-dimensional Jordan block that has $n$ degenerate eigenvalue but only one independent eigenvector, which exists only in non-Hermitian systems that involve dissipation processes.

Recently, experimental studies of (EP) in non-Hermitian systems have dramatically increased \cite{El-Ganainy2018,Feng2017}, whose physical realizations range over electrical circuits \cite{Stehmann2003}, Bose-Einstein condensates \cite{Brinker2015}, optical lasers \cite{Liertzer2012}, dielectric microcavities \cite{Cao2015} and coupled waveguides \cite{Doppler2016}. These experiments are mostly motivated by the counter-intuitive physics induced by the branch-singularity at the EP, owing to which various phase transitions can be observed when the characteristic parameter crosses some certain critical value. Typically, the Parity-Time ($\mathcal{PT}$) symmetry breaking is always associated with an EP  \cite{Bender1998}.

The EP-induced sharp phase transition has inspired many advanced photonic applications \cite{Okolowicz2009,Doppler2016}. With coupled waveguides or resonators, EPs are generated by properly tuning their frequencies and coupling strengths. Using the simplest EP$_2$, one can realize nonreciprocal transmission \cite{Peng2014}, Loss-induced suppression and revival of lasing \cite{Peng2016} in coupled high-Q toroid micro-resonators, and energy transfer between normal modes in coupled waveguide systems \cite{Xu2016,Doppler2016}. With higher-order EPs, the induced phase transition are much more complicated and hence are physically more interesting \cite{Ding2016,Othman2017}. Recent experiments have demonstrated enhanced spontaneous emission \cite{Lin2016} and much higher sensitivity of optical sensors with the an EP$_3$ realized by three resonators \cite{Schnabel2017,Hodaei2017}.

With rapidly developing fabrication technologies, more precise and tunable couplings between multiple resonators have become possible. For example, the system of three coupled ultrahigh-Q microtoroid resonators have been experimentally reported, which can be applied to form complex photonic molecules \cite{Yang2017}. In \cite{Li2017}, mode hybridization was observed in similar photonic molecules consisting of up to six coupled microsphere resonators. Recently, tunable and strong coupling was also achieved with microscale mechanical graphene resonators mechanical resonators \cite{Luo2018}.
From an engineering point of view, such progresses have enabled the system designs for more complex EPs with multiple resonators. In most general cases, this problem can be ascribed to solving a group of nonlinear equations of the coupling constants, which are usually intractable. The existing results are either based on two resonators that is completely solvable, or are obtained under certain symmetry \cite{Jing2017,Schnabel2017,Othman2017}. Under more general circumstances, there are no specific studies on the design of EP.

In this paper, we show that, beyond the two-resonator systems, the EP synthesis in three-resonator systems are fully solvable. Given arbitrary three resonators that have identical resonant frequencies but different loss rates, we can specify all physically admissible EPs and calculate the corresponding coupling strengths for networking the three resonators. In particular, we find a new class of EPs that has not been reported in the literature.

In the remainder of the paper, we will present and pdiscuss these findings in details. Section \ref{Sec:model} introduces the model of coupled-resonator networks, following which Section \ref{Sec:Classification} presents the full classification of all possible EPs that can be hosted in three-resonator networks. In Section \ref{Sec:Transmission}, we illustrate by numerical examples how the EP affect the transmission spectra of external fields coupled to these networks. Finally, Section \ref{Sec:five} concludes our work and provides perspectives for future studies.

\section{The theoretical model of general coupled-resonator networks}\label{Sec:model}
\begin{figure}
\centering
\includegraphics[width=0.8\columnwidth]{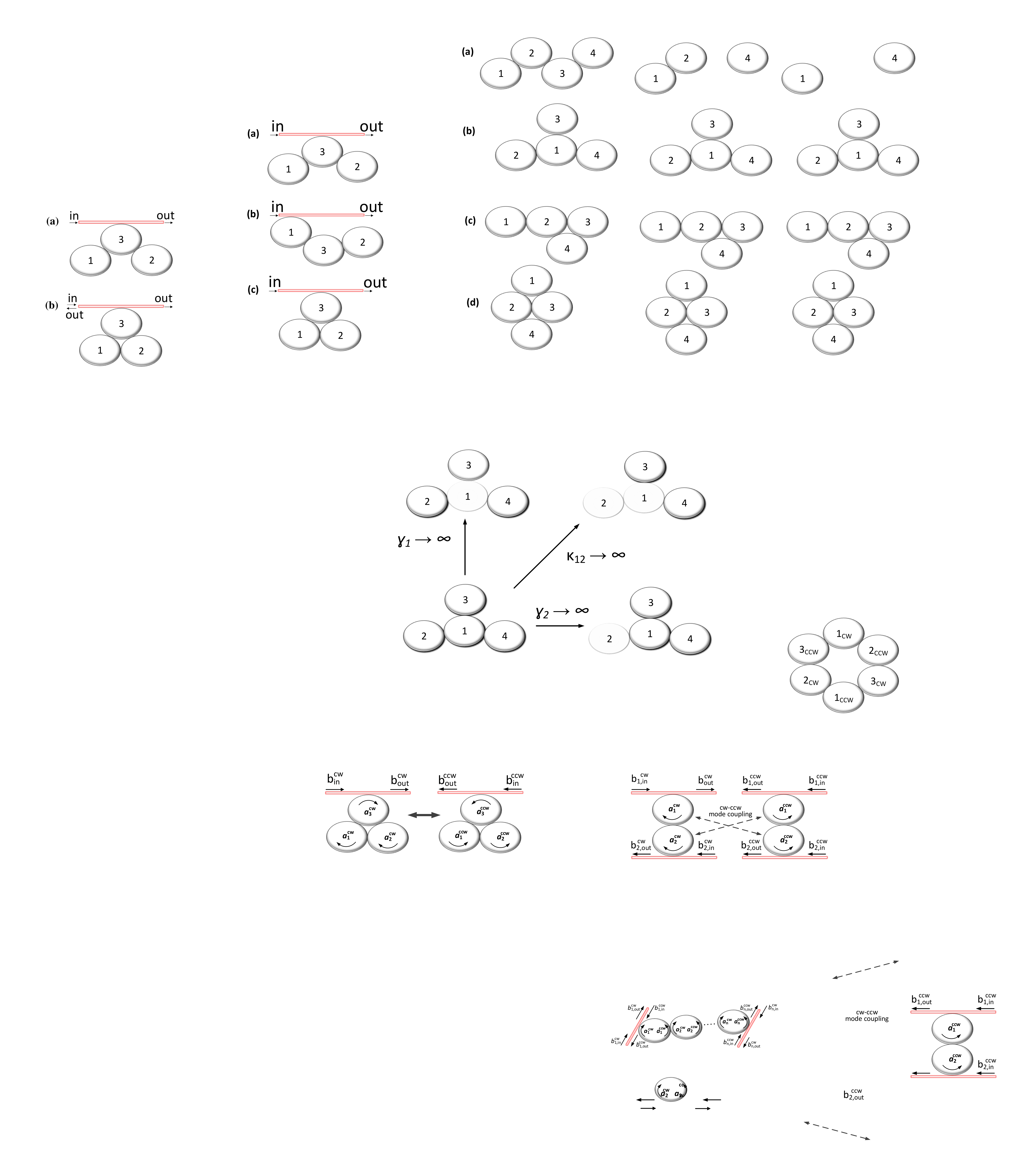}
\caption{The clockwise ($a_k^{\rm cw}$) and counter-clockwise ($a_k^{\rm ccw}$) chiral modes in a waveguide-coupled resonator network. The cw (ccw) modes interact only with ccw (cw) modes with opposite chirality in their neighboring resonators.}\label{fig:mode}
\end{figure}

Consider a network of $n$ resonators (see Fig.~\ref{fig:mode} for the example of a linear chain of resonator network) that are all resonant with each other. In the rotating reference frame, these frequencies can all be set to zero. Let $a_k^{\rm cw}(t)$ and $a_k^{\rm ccw}(t)$ be the clockwise (cw) and counter-clockwise (ccw) chiral modes in the $k$-th resonator, and $b_{k,in/out}^{\rm cw/ccw}$ are the input/output fields coupled to the cw/ccw modes in the $k$-th resonator. As can be seen in Fig.~\ref{fig:mode}, optical modes in two coupled resonators can have interactions only when their chiralities are opposite. Hence, the network dynamics can be described by the following linear differential equations:
\begin{eqnarray*}
\dot x^{\rm cw/ccw} &=& \Gamma x^{\rm cw/ccw} + {\rm K}x^{\rm ccw/cw}+ Bb^{\rm cw/ccw}_{in},\label{eq:2a} \\
b^{\rm cw/ccw}_{out} & = &  b^{\rm cw/ccw}_{in} -B x^{\rm cw/ccw},\label{eq:2b}
\end{eqnarray*}
where $B={\rm diag}\left(\sqrt{2\kappa_1},\cdots,\sqrt{2\kappa_n}\right)$,
\begin{eqnarray*}
x^{\rm cw/ccw}(t)&=&[a_1^{\rm cw/ccw},\cdots,a_n^{\rm cw/ccw}]^\top, \\
b^{\rm cw/ccw}_{in/out}(t)&=&[b_{1,in/out}^{\rm cw/ccw},\cdots,b_{n,in/out}^{\rm cw/ccw}]^\top
\end{eqnarray*}
 and
\begin{eqnarray*}\label{}
\Gamma = \left(
	\begin{array}{ccc}
		\gamma_1 &  & \\
		 & \ddots& \\
		 &   & \gamma_n
	\end{array}
	\right),\quad	{\rm K} = \left(
	\begin{array}{ccc}
		0 &  \cdots&\kappa_{1n} \\
		\vdots & \ddots&\vdots \\
		\kappa_{1n} &  \cdots & 0
	\end{array}
	\right).
\end{eqnarray*}
In these matrices, $\gamma_k$ is the energy exchange rate of the $k$-th resonator with its environment, which is negative or positive when the resonator has loss (called passive) or gain (called active). The coupling strength between the $i$-th and the $j$-th resonators is represented by $\kappa_{ij}$, and $\kappa_i$ is the coupling strength of the $i$-th resonator to the field fed into it.

To facilitate the analysis, the above coupled-mode equations can be decomposed into the following non-interacting parts:
\begin{eqnarray}
\dot x^{\pm} &=& (\Gamma\pm\imath {\rm K})x^{\pm} + Bb^{\pm}_{in},\label{eq:quad1} \\
b_{out}^{\pm}  & = &  b_{in}^{\pm}-B x^{\pm},\label{eq:quad2}
\end{eqnarray}
under the quadrature representation:
\begin{eqnarray}
x^{\pm} &=& \frac{1}{\sqrt{2}}\left(x^{\rm cw}\pm x^{\rm ccw}\right), \label{eq:transform1}\\
b_{in,out}^{\pm} &=& \frac{1}{\sqrt{2}}\left(b_{in,out}^{\rm cw}\pm b_{in,out}^{\rm ccw}\right).\label{eq:transform2}
\end{eqnarray}

The network is said to possess an EP$_n$ if and only if the matrices $\Gamma\pm \imath {\rm K}$ have an $n$-dimensional Jordan block. Since the two matrices are complex conjugate with each other, it is sufficient to study only one of them, say
\begin{equation}\label{Eq:A}
A =\Gamma+\imath {\rm K}=\left(
	\begin{array}{ccc}
		\gamma_1 &  \cdots& \imath\kappa_{1n} \\
		\vdots & \ddots&\vdots \\
		\imath\kappa_{1n} &  \cdots & \gamma_n
	\end{array}
	\right),
\end{equation}
and any conclusion about $A$ can be directly extended to $A^*=\Gamma-\imath {\rm K}$.

Suppose that $\sigma$ is the $n$-fold degenerate eigenvalue corresponding to an EP$_n$, and $\vec{v}=[v_1,\cdots,v_n]^T$ is the corresponding eigenvector. Then, the real and imaginary parts of $\sigma$ represent the the loss (or gain) rate and resonant frequency of the eigenmode associated with this EP$_n$, and each $|v_k|^2$ represents the portion of the mode's stationary power allocated to the $k$-th resonator.

It is easy to derive from the above model the input-output relation between the external fields coupled to the network. Let $\hat{x}(s)$ be the Laplace transform of $x(t)$. By Laplace transforming Eqs.~(\ref{eq:quad1}) and (\ref{eq:quad2}), we can obtain the transfer function from the quadrature inputs to the quadrature outputs, as follows:
\begin{equation}
\hat b_{out}^\pm(s) = \left[\mathbb{I}_n-B(s\mathbb{I}_n-\Gamma\mp\imath {\rm K})^{-1}B\right]\hat b_{in}^\pm(s),
\end{equation}
where it is assumed that the resonators are initially in vaccum; and $\mathbb{I}_n$ is the identity matrix. Using the transformation (\ref{eq:transform1}) and (\ref{eq:transform2}), we can obtain the transfer functions in the propagating-wave representation:
\begin{equation}\label{eq:IO}
\left[
  \begin{array}{c}
    \hat b_{out}^{\rm cw}(s) \\
    \hat b_{out}^{\rm ccw}(s) \\
  \end{array}
\right] = \left[
            \begin{array}{cc}
              T(s) & R(s) \\
              R(s) & T(s) \\
            \end{array}
          \right]
\left[
  \begin{array}{c}
    \hat b_{in}^{\rm cw}(s) \\
    \hat b_{in}^{\rm ccw}(s) \\
  \end{array}
\right],
\end{equation}
where the matrix transfer functions are defined by
\begin{eqnarray*}
  T(s) &=& \mathbb{I}_n-\frac{1}{2}B\left[(s\mathbb{I}_n-A)^{-1}+(s\mathbb{I}_n-A^*)^{-1}\right]B \label{eq:IO1}\\
  R(s) &=& -\frac{1}{2}B\left[(s\mathbb{I}_n-A)^{-1}-(s\mathbb{I}_n-A^*)^{-1}\right]B.\label{eq:IO2}
\end{eqnarray*}
Their $(i,j)$-th entry, say $T_{ij}(s)$ or $R_{ij}(s)$, describes the transmission or reflection property from the $j$-th port to the $i$-th port, i.e., the ratio between the cw and ccw fields output from the $i$-th resonator and the cw field input to the $j$-th resonator. The transmission or reflection spectra can be obtained by simply setting $s=\imath\omega$, as long as the entire network system is stable, i.e., when all eigenvalues of $A$ have negative real parts.

In principle, the input field fed into any resonator can come out from any other port and go into either the clockwise or the counter-clockwise output fields. However, some transmissions can be inhibited under certain topologies. For example, if there are no circular couplings of odd number of resonators [e.g., the circularly coupled three resonators shown in Fig.~\ref{fig:chain}(c)], the input field through the $i$-th port will be fed forward into the network without being scattered back into the same input port, i.e., all reflection transfer function $R_{ii}(s)$ vanishes. This property can be utilized in the synthesis of resonator networks when backscattering is unwanted.

\section{EP Classification in three-resonator networks}\label{Sec:Classification}
\begin{figure}[H]
\begin{center}
\includegraphics[width=0.5\columnwidth]{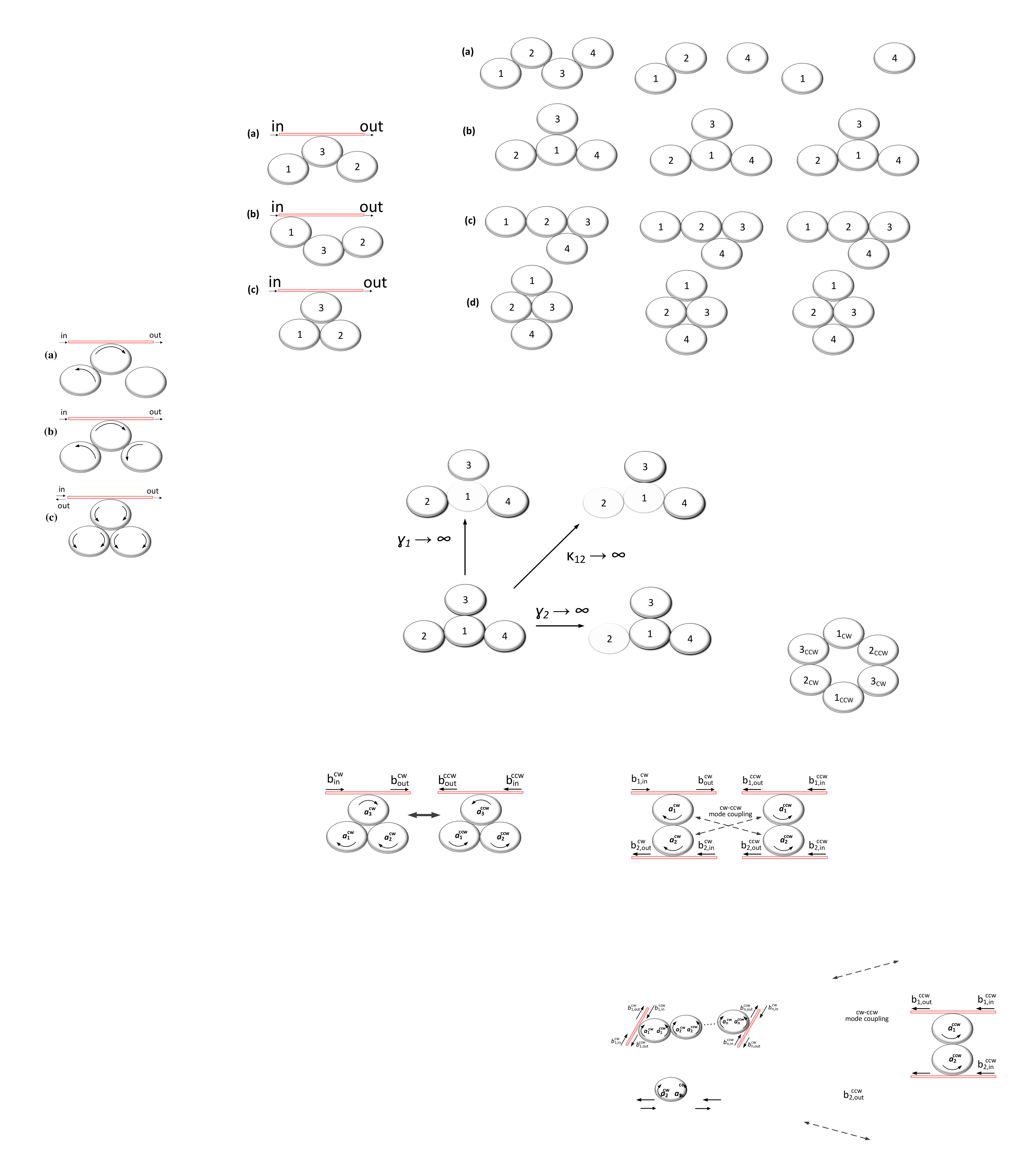}
\end{center}
\caption{\small  Schematic diagram for possible topologies of three-resonator networks with input/output fields coupled to the middle resonator. (a) one resonator is isolated from the other two coupled resonators; (b) three linearly coupled resonators; (c) three circularly coupled resonators. The fields travels unidirectionally in (a) and (b), while in (c) both chiral modes coexist.}\label{fig:chain}
\end{figure}

In this paper, we are concerned with three-resonator networks. As shown in Fig.~\ref{fig:chain}, there are only three possible network topologies in which an EP can possibly exist. Figure \ref{fig:chain}(a) shows an actual two-resonator network that has been well studied in the literature. What we are mainly concerned with are the non-trivial three-resonator networks shown in Figs.~\ref{fig:chain}(b) and \ref{fig:chain}(c), in which the three resonators are either linearly or circularly coupled.

Let the loss (gain) rates of the three resonators be $\gamma_1$, $\gamma_2$ and $\gamma_3$. Without loss of generality, we assume throughout this paper that $\gamma_1<\gamma_2<\gamma_3$ and $\gamma_1+\gamma_2+\gamma_3=0$. We make the latter assumption because the coalesce of eigenvalues and eigenvectors is not changed after subtracting $A$ by $\gamma_0 \mathbb{I}_3$, where $\gamma_0=\frac{1}{3}(\gamma_1+\gamma_2+\gamma_3)$ is the average loss rate, and therefore it is sufficient to consider only traceless $A$. Such simplification may greatly facilitate our following analysis. The resulting balanced loss-gain networks are also physically interesting because the entire network has no net energy exchange with its environment.

Before expanding our discussion, we introduce the following constant
\begin{equation}\label{eq:Delta}
\Delta^2 = \frac{1}{18}\left[(\gamma_1-\gamma_2)^2+(\gamma_2-\gamma_3)^2+(\gamma_3-\gamma_1)^2\right], \end{equation}
that characterizes the non-Hermitiancy of the network. When $\Delta=0$, $A$ is diagonalizable and hence can never been associated with an EP. In addition, we define
\begin{equation}
\kappa^2 = \frac{1}{3}(\kappa_{12}^2 + \kappa_{23}^2 + \kappa_{31}^2)
\end{equation}
as the average coupling strength. In Appendix A, we prove that, if the network has an EP$_2$ eigenvalue at $\sigma$, it must satisfy the following simple relation:
\begin{equation} \label{eq:a1}
\sigma^2=\Delta^2-\kappa^2.
\end{equation}
When $\kappa<\Delta$, we say that the network is weakly coupled and in this regime $\sigma^2>0$, i.e., $\sigma$ must be a real number. When $\kappa>\Delta$, we say that the network is strongly coupled, and in this regime $\sigma^2<0$, which implies that $\sigma$ must be an imaginary number.

\begin{figure}
\centering{
\includegraphics[width=0.9\columnwidth]{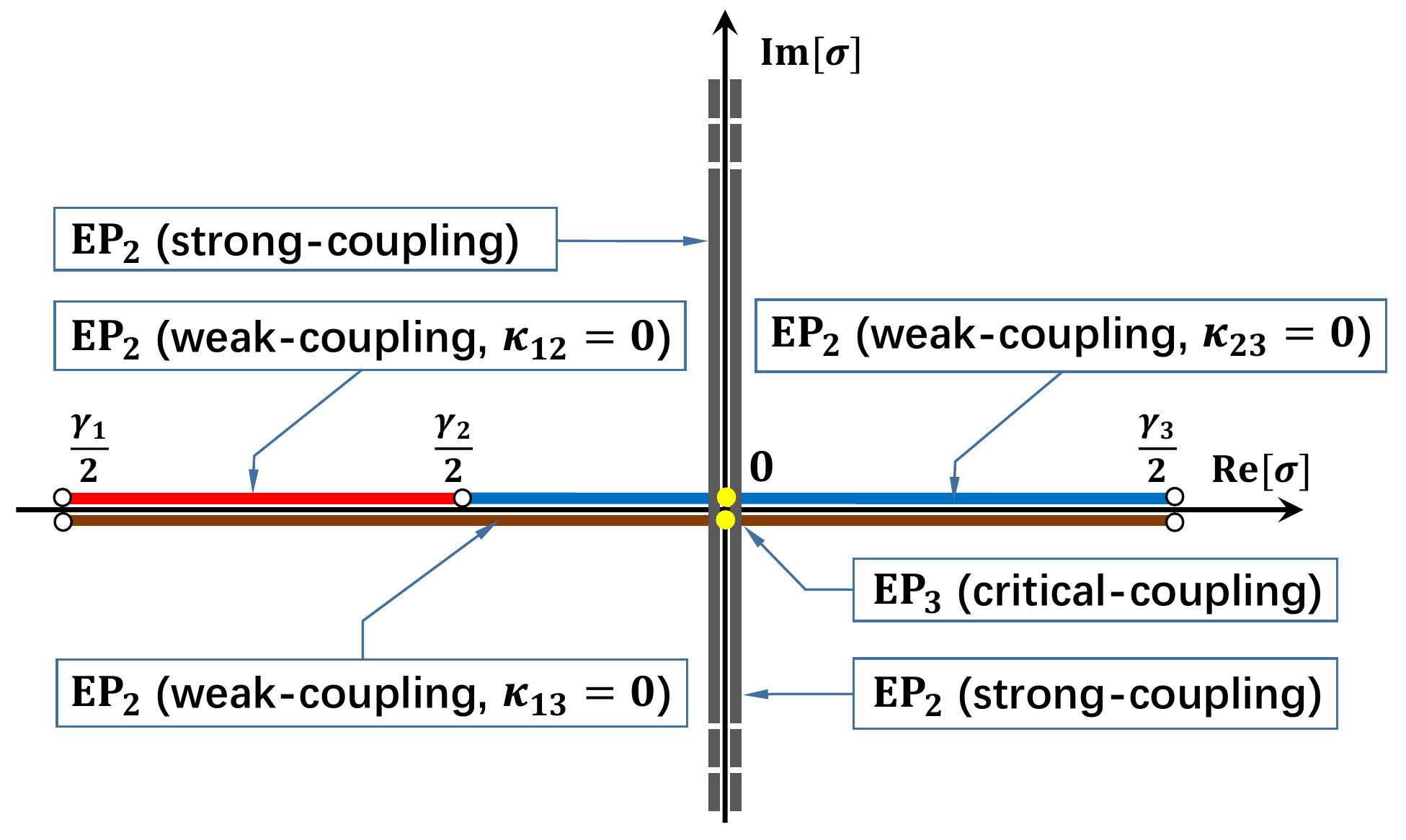}}
\caption{\small Schematic diagram for the distribution of degenerate eigenvalues and their associated twin EPs in a given three-resonator network ($\gamma_1\leq\gamma_2\leq\gamma_3$). Each point in the brown (red, blue) horizontal line segments represents an EP$_2$ (with eigenvalue being real numbers) in the weak-coupling regime, which are realized by linear networks with $\kappa_{13}=0$ ($\kappa_{12}=0$, $\kappa_{23}=0$), respectively. The two vertial lines represent the twin EPs (with eigenvalues being imaginary numbers) in the strong coulping regime, which are realized by circular networks. The two yellow points at the intersection of weak-coupling and strong-coupling EPs represent the twin EP$_3$ realizations in the critical coupling regime.}\label{fig:EPdistribution}
\end{figure}

In the appendices, we prove that for every admissible EP eigenvalue $\sigma$, there exist two and only two resonator-network realizations, and we call them {\it twin} EPs associated with $\sigma$. The distribution of all these EPs is shown in Fig.~\ref{fig:EPdistribution}. Most of them are EP$_2$, and EP$_3$ only appears at the intersection of weak and strong coupling regimes, which we call critical-coupling regime.

\subsection{The weak coupling regime}

\begin{figure}
\centering
{\includegraphics[width=0.8\columnwidth]{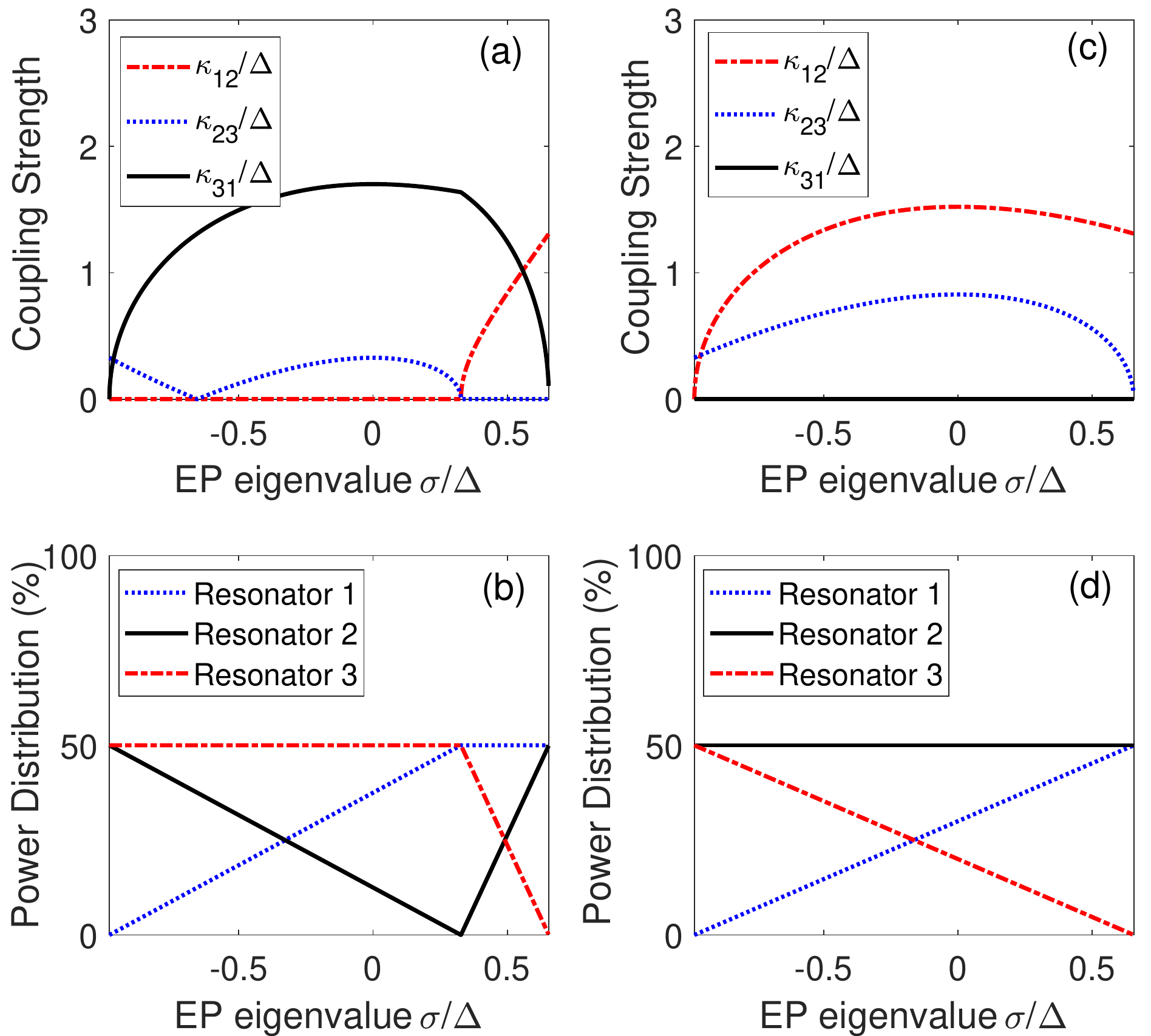}
	}	
\caption{\small The twin weak-coupling EP realizations [(a)-(b) for the first and (c)-(d) for the second] with resonators $\gamma_1/\Delta=-1.9640$, $\gamma_2/\Delta=0.6547$ and $\gamma_3/\Delta=1.3093$. The plots (a) and (c) show the coupling strengths for each pair of twin EPs, while (b) and (d) are the power distribution in the corresponding EP eigenmodes.}\label{fig:kappaC}
\end{figure}

As is proven in Appendix A, EPs in the weak-coupling regime are all realized by linear networks, and the corresponding degenerate eigenvalues are real numbers ranging from $\gamma_1/2$ to $\gamma_3/2$. The corresponding eigenmodes are on-resonance with the bare resonators, but are non-conservative (i.e., being passive or active at the rate of $\sigma$).

For each admissible EP eigenvalue $\sigma$, we can find two and only two EP$_2$ linear network realizations. For example, if $\sigma\in(\gamma_1/2,\gamma_{2}/2)$, one realization has the third resonator in the middle with the first and second resonators uncoupled (i.e., $\kappa_{12}=0$, see the upper red line segment on the left in Fig.~\ref{fig:EPdistribution}), and the other resonator has the second resonator in the middle with the first and the third resonators uncoupled (i.e., $\kappa_{31}=0$, see the lower black line segment in Fig.~\ref{fig:EPdistribution}).

These twin EPs are mostly realized by linearly coupled networks shown in Fig.~\ref{fig:chain}(b) except when $\sigma= \gamma_1/2$, $\gamma_1/2$ or $\gamma_3/2$, in which cases one resonator is decoupled from the other two, resulting in actual two-resonator networks. For example, when $\sigma=\gamma_2/2$, both $\kappa_{12}$ and $\kappa_{23}$ vanish according to Eqs.~(\ref{eq:EP2c1}) and (\ref{eq:EP2c2}), which leads to a network shown in Fig.~\ref{fig:chain}(a) where the second resonator is disconnected with the other two. Such EP is not the focus of this paper. There are totally five such networks as indicated by the circles in Fig.~\ref{fig:EPdistribution}. In addition, if it happens that $-\gamma_k\in(\gamma_1/2, \gamma_3/2)$ for some $1\leq k\leq 3$, a two-resonator network also exists with $\sigma=-\gamma_k$.

To see how the coupling strengths at these twin EPsvary with the EP eigenvalues. We choose $\gamma_1=-3$ (a.u.), $\gamma_2=1$(a.u.) and $\gamma_2=2$ (a.u.), and normalize them by the non-Hermitiancy measure $\Delta=\sqrt{7/3}$ (a.u.) that is defined by (\ref{eq:Delta}), which results in
\begin{equation}\label{eq:para}
\frac{\gamma_1}{\Delta}=-1.9640, \frac{\gamma_2}{\Delta}=0.6547,\frac{\gamma_3}{\Delta}=1.3093.
\end{equation}
Then, for each admissible $\sigma\in[\gamma_1/2,\gamma_3/2]$, we calculate the coupling strengths (normailzed by $\Delta$) with these resonators according to the formulas (\ref{eq:EP2c1}) and (\ref{eq:EP2c2}) derived in Appendix A, as well as the power distribution in the eigenmode. The results for the two EPs are depicted in Fig.~\ref{fig:kappaC}(a)-(b) and Fig.~\ref{fig:kappaC}(c)-(d), respectively.

It can be seen that for each EP$_2$ at least one coupling constant is zero, which is consistent with the conclusion that weak-coupling EPs are all realized by linear networks. At special values, one resonator is isolated from the other two, leaving only one coupling constant nonzero. Associated with the given three resonators, there are totally six cases [4 in Fig.~\ref{fig:chain}(a) and 2 in iFig.~\ref{fig:chain}(c)], which correspond to $\sigma=\gamma_1/2,\gamma_1/2,\gamma_1/2,-\gamma_2$, respectively.

When the network is operated at the EP mode, the stationary power distribution in each resonator (corresponding to the square norm $|v_1|^2$, $|v_2|^2$ and $|v_3|^2$ of the entries of the eigenvector) is shown in Figs.~\ref{fig:kappaC}(c) and \ref{fig:kappaC}(d) for the twin EPs. We observe that the resonator in the middle always occupies half of the total power, while the other two side resonators share the rest half. When the EP eigenvalue $\sigma$ approaches to some $\gamma_k/2$ ($k=1,2,3$), the power in the $k$-th resonator tends to be pushed out. In the limit $\sigma=\gamma_k/2$, the $k$-th resonator is completely isolated and hence no power remains in this resonator.



\subsection{The strong-coupling regime}

\begin{figure}
\centering
{\includegraphics[width=0.8\columnwidth]{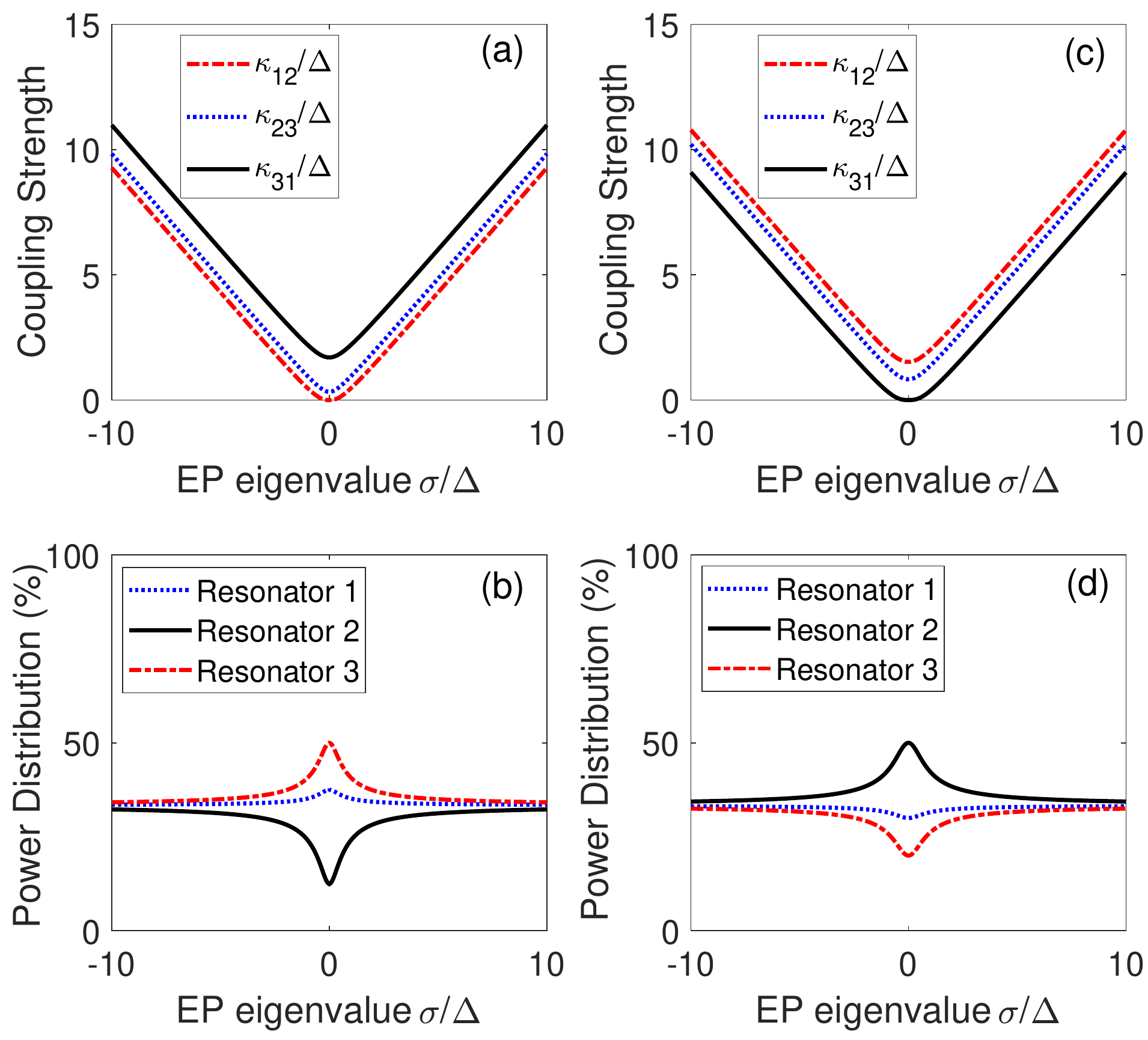}
	}	
\caption{\small The twin strong-coupling EP realizations [(a)-(b) for the first and (c)-(d) for the second] with resonators $\gamma_1/\Delta=-1.9640$, $\gamma_2/\Delta=0.6547$ and $\gamma_3/\Delta=1.3093$. The plots (a) and (c) are the coupling strengths for each admissible EP, while (b) and (d) are the power distribution in the corresponding EP eigenmodes.}\label{fig:kappaR}
\end{figure}

In the strong coupling regime $\kappa>\Delta$, the EP eigenvalue is always purely imaginary, say $\imath \sigma$ with $\sigma\in\mathbb{R}$. The corresponding eigenmode is energy conservative and oscillates with frequency $\sigma$.

The EP analysis in the strong-coupling regime is more complicated than that in the weak-coupling regime. In Appendix B, we prove that every $ \imath\sigma\in\mathbb{R}$ is also associated with exactly two EP$_2$, as shown in Fig.~\ref{fig:EPdistribution} by the vertical lines along the imaginary axis. Such twin EP$_2$ are all realized by circular networks.

The coupling strengths for these twin EPs in the strong-coupling regime can be numerically calculated by solving the roots of a cubic equation (see Appendix B.1 for the procedure). Only when the network is symmetric (i.e., two resonators have identical loss rates), there exists an analytic formula (see Appendix B.2). Using these methods, we calculate and plot in Fig.~\ref{fig:kappaR} the dependence of the coupling strengths of the twin EPs with respect to $\sigma$, in which the same sets of resonator parameters (\ref{eq:para}) are used. Except at $\sigma=0$, all coupling constants are nonzero, which indicates that these EPs are realized by circular networks. Moreover, when the magnitude of the desired $\sigma$ grows, all coupling strengths increase as well, and they tend to rise linearly when $\sigma$ is large.

The coalescing eigenvectors at these EPs also have to be numerically calculated, and the corresponding stationary power distribution in Figs.~\ref{fig:kappaR}(b) and \ref{fig:kappaR}(d) for the twin EPs. It can be seen that the strong coupling between two resonators tends to retain more power in them and leave less power to the rest resonator. For example, the strongest coupling in Fig.~\ref{fig:kappaR}(a) is  $\kappa_{31}$ and, correspondingly, the power allocated to the second resonator is the least, as shown in Fig.~\ref{fig:kappaR}(b).  Moreover, the power is more evenly distributed in the three resonators, which is different from EPs in the weak-coupling regime. When $\sigma$ goes to infinity, the power tends to distribute uniformly in the three resonators.

\subsection{The critical coupling regime}\label{Sec:Model-EP3}

At the intersection of weak- and strong-coupling regimes, the case $\kappa=\Delta$ is special because only under this condition can an EP$_3$ be realized. Similarly, we have also twin EP$_3$ as shown by the two yellow circles at $\sigma=0$ in Fig.~\ref{fig:EPdistribution}, and they are realized by linearly coupled networks. The parametrization of the associated $A$ can be directly obtained from Eqs.~(\ref{eq:EP2c1}) and (\ref{eq:EP2c2}) by setting $\sigma=0$. For example, if $\sigma=0<\gamma_2/2$, we have one EP$_3$ with $\kappa_{31}=0$ and
\begin{equation}\label{eq:EP3a}
\kappa_{12}= \gamma_1\sqrt{\frac{\gamma_1}{\gamma_1-\gamma_3}},
\quad\kappa_{23}= \gamma_3\sqrt{\frac{\gamma_3}{\gamma_3-\gamma_1}},
\end{equation}
and its twin EP$_3$ has $\kappa_{23}=0$ and
\begin{equation}\label{eq:EP3b}
\kappa_{12}= \gamma_2\sqrt{\frac{\gamma_2}{\gamma_2-\gamma_3}},
\quad\kappa_{31}= \gamma_3\sqrt{\frac{\gamma_3}{\gamma_3-\gamma_2}}.
\end{equation}

Note that there is an exceptional case in which one of the two EPs is an EP$_2$ but not EP$_3$, i.e, the eigenvalues are three-fold degenerate, but only two eigenvectors becomes parallel. This happens when the networks when $\gamma_1=-\gamma_3$ and $\sigma=0$).
In this case, $\kappa_{12}=\kappa_{23}=0$ [using Eqs.~(\ref{eq:EP3a}) and (\ref{eq:EP3b})] for one of the two EPs, which becomes a two-resonator network with the second resonator being isolated. The coupling strengths associated with its twin EP are $\kappa_{31}=0$ and $\kappa_{12}\neq 0\neq \kappa_{23}$, which can be verified to still be an EP$_3$. This kind of EP$_3$ possesses $\mathcal{PT}$ symmetry and has been adopted in \cite{Hodaei2017} for sensing applications. Our analysis shows that more general EP$_3$ can be synthesized with any three mutally resonant resonators, which is in practice useful when $\mathcal{PT}$-ymmetric networks are not easy to realize.

\section{Transmission properties near exceptional points}\label{Sec:Transmission}

One of the most intriguing applications of EPs is sensing, because the measurement system is very sensitive to the change of parameter when being operated near an EP, and its sensitivity increases with the order of the EP. In this section,  we will demonstrate such properties with the above classified EPs from the perspective of sensing applications.

\subsection{EP$_2$ in the weak-coupling regime}	

\begin{table}
\begin{tabular}{c|c|c|c|c|c|c}
  \hline
\multicolumn{3}{c|}{Loss rates}  &  \multicolumn{3}{c|}{Coupling} & \multirow{2}*{Type}  \\
\cline{1-3}\cline{4-6}
$\gamma_1/\Delta$ & $\gamma_2/\Delta$ & $\gamma_3/\Delta$ &  $\kappa_{12}/\Delta$ &  $\kappa_{23}/\Delta$ & $\kappa_{31}/\Delta$ &  \\
  \hline
  \multirow{2}*{-1.3093} &   \multirow{2}*{-0.6547} &   \multirow{2}*{1.9640}  & 0.8281 & 1.5213 & 0 & EP$_2$ \\
  \cline{4-7}
   & & & 0.3273 & 0 & 1.7008 & EP$_2$ \\
  \hline
\end{tabular}
\caption{\small  The coupling constants of the twin EP$_2$ associated with the EP eigenvalue at $\sigma/\Delta=0.1$, which are realized by linearly coupled three-resonator networks in the weak-coupling regime. }\label{Tab1}
\end{table}

Consider the same example with parameters given by (\ref{eq:para}). We assign the degenerate EP$_2$ eigenvalue at $\sigma/\Delta=0.1$ and calculte the corresponding coupling constants for two distinct realizations according to Eqs.~(\ref{eq:EP2c1}) and (\ref{eq:EP2c1}). The results are listed in Table \ref{Tab1}.

We then perturb the loss rate $\gamma_3$ by a small quantity $\epsilon$ that ranges from $-0.01\Delta$ to $0.01\Delta$, and observe how the eigenmodes and transmission spectra (through the waveguide coupled to the third resonator) of the network vary with the perturbation. The transmission spectrum is calculated from the input-output transfer function (\ref{eq:IO1}) derived in Section \ref{Sec:model}. Since the balanced loss-gain system is physically instable (with eigenvalues at $\sigma=0.1\Delta$ and $-0.2\Delta$), we shift the designed $A$ to $A-(\sigma+\gamma) \mathbb{I}_3$ when simulating the transmission spectra, where $\gamma=10^{-3}\Delta$ is introduced to guarantee the stability of the entire network. This does not change the nature of the EP structure and the corresponding coupling strengths are still the same.

\begin{figure}
\begin{center}
\includegraphics[width=0.8\columnwidth]{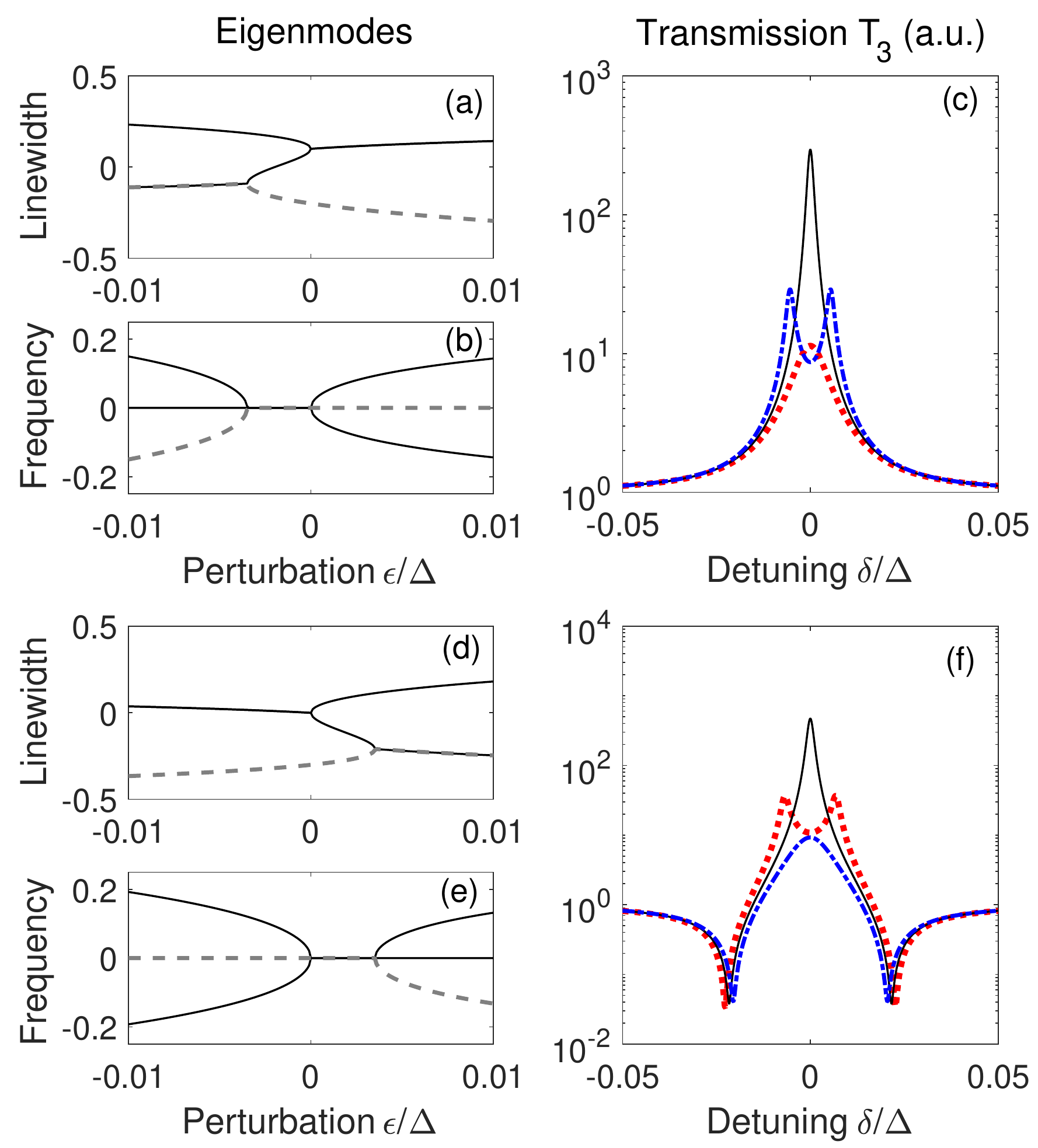}
\end{center}	
\caption{\small The eigenmodes (left) and the transmission spectra (right) near the twin weak-coupling EP$_2$ [corresponding to (a)-(c) and (d)-(f)] designed at $\sigma/\Delta=0.1$. The transmission spectra at the EP$_2$ are the black solid curves, while the red dotted (blue dash-dotted) curves are the transmission spectra when $\gamma_3$ is perturbed by $\epsilon/\Delta=-10^{-5}$ ($\epsilon/\Delta=10^{-5}$). The loss (gain) rates of the resonators are chosen as $\gamma_1/\Delta=-1.3093$, $\gamma_2/\Delta=-0.6547$ and $\gamma_3/\Delta=1.9640$.}\label{fig:EP2c}
\end{figure}

The variance of the eigenmodes with the perturbation parameter $\epsilon$ are shown in the plots in left column of Fig.~\ref{fig:EP2c}, while the transmission spectra are plotted in the right column. The system experiences a phase transition near the EP. Take the first EP [Figs.~\ref{fig:EP2c}(a)-(c)] for example, the EP is assigned at $\epsilon/\Delta=0$. When $\epsilon<0$, the three eigenmodes near the EP have identical imaginary parts (i.e., they are all resonant with the bare resonators) but different real parts (i.e., different linewidths), which is manifested by the only one peak appearing in the transmission spectra (the blue curve in the plot to the right). After crossing the EP at $\epsilon=0$, two of the three eigenmodes (see the solid curves) have identical real parts but different imaginary parts, which leads to the mode splitting shown by the separated peaks (red curve) with identical widths. This is a $\mathcal{PT}$-symmetry-like phase in which the network behaves like a Hermitian system, which is broken when $\epsilon<0$. The second EP$_2$ (the lower plots) exhibits an opposite phase transition where the $\mathcal{PT}$-symmetry-like phase appears when $\epsilon<0$.

In both transmission spectra, the parameter $\gamma_3$ is perturbed by $\epsilon/\Delta=\pm 10^{-5}$, and the mode splittings observed in Figs.~\ref{fig:EP2c}(c) and \ref{fig:EP2c}(f) are $10^{-3}\sim10^{-2}\Delta$. This is consistent with the expectation that mode splitting is proportional to $\epsilon^{1/2}$ at an EP$_2$, which has been used for improving the sensing sensitivity in various applications \cite{El-Ganainy2018}.

\subsection{EP$_2$ in the strong-coupling regime}
\begin{table}
\begin{tabular}{c|c|c|c|c|c|c}
  \hline
\multicolumn{3}{c|}{Loss rates}  &  \multicolumn{3}{c|}{Coupling} & \multirow{2}*{Type}  \\
\cline{1-3}\cline{4-6}
$\gamma_1/\Delta$ & $\gamma_2/\Delta$ & $\gamma_3/\Delta$   & $\kappa_{12}/\Delta$ &  $\kappa_{23}/\Delta$ & $\kappa_{31}/\Delta$ &  \\
  \hline
  \multirow{2}*{-1.3093} &   \multirow{2}*{-0.6547} &   \multirow{2}*{1.9640}  & 0.3386 & 0.0017 & 1.7074 & EP$_2$ \\
  \cline{4-7}
   & & & 0.8353 & 1.5272 & 0.0008 & EP$_2$ \\
  \hline
\end{tabular}
\caption{\small   The coupling constants of the twin EP$_2$ associated with the EP eigenvalue at $\sigma/\Delta=0.1\imath$, which are realized by linearly coupled three-resonator networks in the strong-coupling regime. }\label{Tab2}
\end{table}

Using the same set of resonators as above, we study the twin EP$_2$ in the strong coupling regime. The degenerate EP$_2$ eigenvalue is chosen at $\sigma/\Delta=0.1\imath$, i.e., the corresponding EP eigenmode is detuned from the resonatant frequency by $0.1\Delta$, and is purely oscillating with neither loss nor gain. The coupling constants are calculated for the twin EPs and are listed in Table \ref{Tab2}.

Similar to the simulations in weak-coupling regime, we perturb the loss rate $\gamma_3$ by some small $\epsilon$ and observe how it affects the eigenmodes and the transmission spectrum. In the simulation of transmission spectra, we also slightly shift the designed $A$ to $A-\gamma \mathbb{I}_3$, where $\gamma=10^{-3}\Delta$ is introduced to guarantee the stability of the entire network.

\begin{figure}
\centering
{\includegraphics[width=0.8\columnwidth]{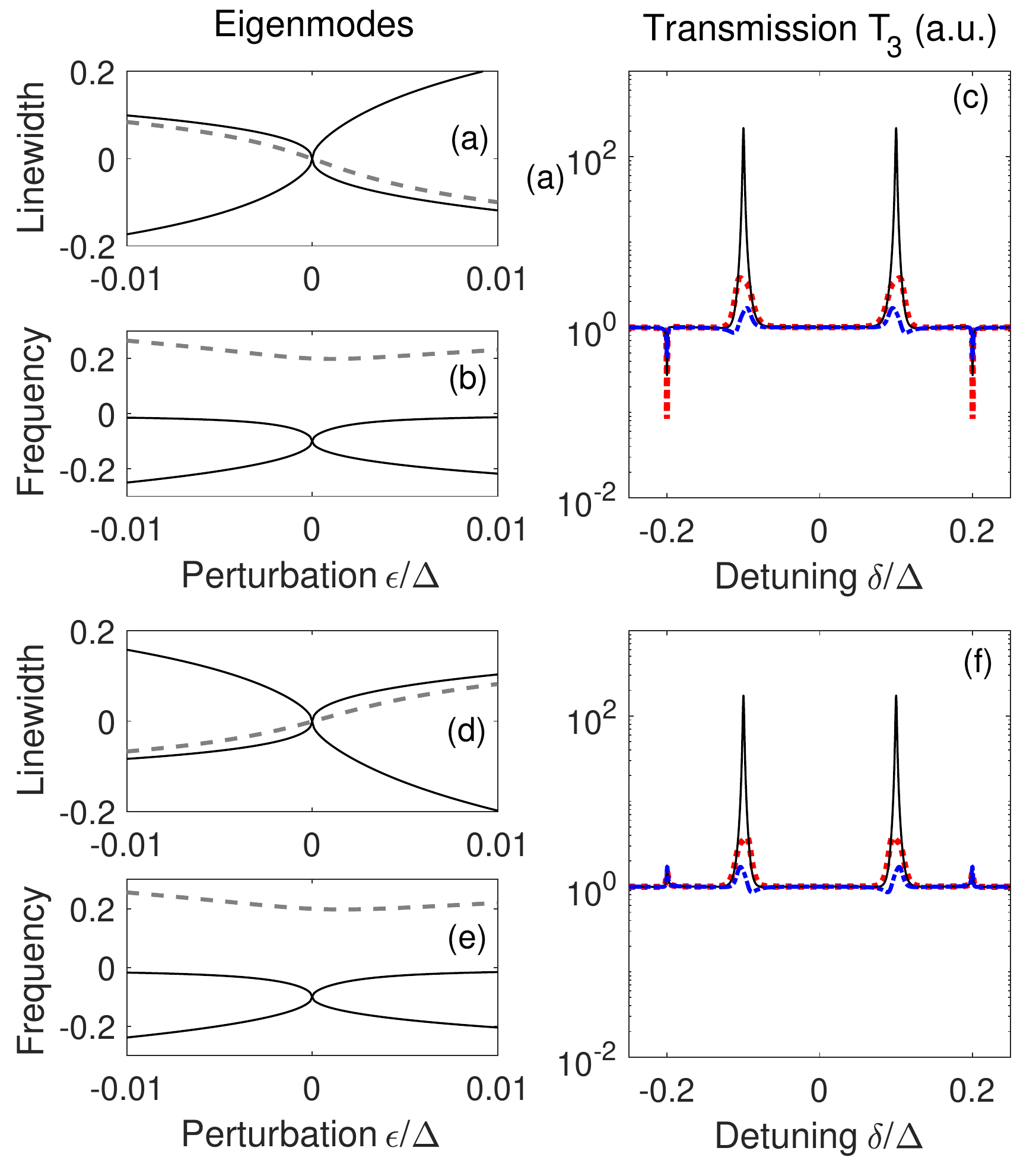}
	}	
\caption{\small The eigenmodes (left) and the transmission spectra (right) near the twin strong-coupling EP$_2$ [corresponding to (a)-(c) and (d)-(f)] designed at $\sigma/\Delta=0.1\imath$. The transmission spectra at the EPs are shown by black solid curves, while the red dotted (blue dash-dotted) curves are the spectra when $\gamma_3$ is perturbed by $\epsilon/\Delta=-10^{-5}$ ($\epsilon/\Delta=10^{-5}$). The loss (gain) rates of the resonators are chosen as $\gamma_1/\Delta=-1.3093$, $\gamma_2/\Delta=-0.6547$ and $\gamma_3/\Delta=1.9640$.}\label{fig:EP2r}
\end{figure}

The plots of the eigenmodes and transmission spectra are shown in the left and right columns, respectively, of Fig.~\ref{fig:EP2r}. As expected, we observe resonant peaks at frequencies $0.1\Delta$ (corresponding to the two degenerate EP eigenvalues) and $-0.2\Delta$ (corresponding to the rest eigenvalue). We also observe two additional peaks at symmetric positions $-0.1\Delta$ and $0.2\Delta$, which is contributed by the complex conjugate $A^*$ of $A$ that is also involved in the transmission spectrum [see Eq.~\ref{eq:IO1}]. Note that in the weak-couping regime the transmission spectra contributed by $A$ and $A^*$ completely overlaps with each other.

Owing to the singularity of EP$_2$ at $\pm 0.1\Delta$, the peaks at $\pm 0.1$ are much sharper than those at $\pm 0.2\Delta$. Near the EP, the linewidths and frequencies split on both sides of the EP [see the plots Figs.~\ref{fig:EP2r}(b) and \ref{fig:EP2r}(e)], and thereby the breaking of $\mathcal{PT}$-symmetry in the weak-coupling regime does not appear here. In the transmission spectra, we perturb $\gamma_3$ by $\pm 0.01\Delta$, and observe sharp changes of the peak heights on both sides. The mode splitting can also be seen but is much less discernable.

Because the strong-coupling EP$_2$ is realized by circularly coupled networks, the coexistence of both cw and ccw modes in the resonators may induce backscattering (or reflection) in the coupled waveguide [see the leftwards output field in Fig.~\ref{fig:chain}(c)]. This does not happen for weak-coupling EP$_2$'s owing to the corresponding linear topology. The backscattering can be observed in the reflection spectrum, whose characteristics are similar to those of the transmission and will not be separately discussed here.

\subsection{EP$_3$ and EP$_2$ in the critical coupling regime}
\begin{table}
\begin{tabular}{c|c|c|c|c|c|c}
  \hline
\multicolumn{3}{c|}{Loss rate}   &  \multicolumn{3}{c|}{Coupling strength} & \multirow{2}*{Type}  \\
\cline{1-3}\cline{4-6}
\;$\gamma_1/\Delta$\; & \;$\gamma_2/\Delta$\; & \;$\gamma_3/\Delta$\; &  \;$\kappa_{12}/\Delta$\; &  \;$\kappa_{23}/\Delta$\; & \;$\kappa_{31}/\Delta$\; &  \\
  \hline
  \multirow{2}*{-1.3093} &   \multirow{2}*{-0.6547} &   \multirow{2}*{1.9640} & 0.8281 & 1.5213 & 0 & EP$_3$ \\
   \cline{4-7}
   & & & 0.3273 & 0 & 1.7008 & EP$_3$ \\
\hline
  \multirow{2}*{-1.7321} &   \multirow{2}*{0} &   \multirow{2}*{1.7321} & 1.2247 & 1.2247P & 0 & EP$_3$ \\
   \cline{4-7}
   & &  & 0 & 0 & 1.7321 & EP$_2$ \\
  \hline
\end{tabular}
\caption{\small  The coupling constants of two pairs of twin EPs, which are realized two different sets of resonators in the critical-coupling regime with $\sigma/\Delta=0$. The first three are all EP$_3$. The last EP has a three-fold degenerate eigenvale but two-fold degenerate eigenvector and hence is an EP$_2$.}\label{Tab3}
\end{table}

Using the same set of resonators, we calculate the coupling constants for the twin EP$_3$ at $\sigma=0$, which is both lossless and on-resonance with the bare resonators. They are realized by linear networks with $\kappa_{12}=0$ and $\kappa_{31}=0$ (see the upper two rows in Table \ref{Tab3}), respectively.

\begin{figure}
\centering
\includegraphics[width=0.8\columnwidth]{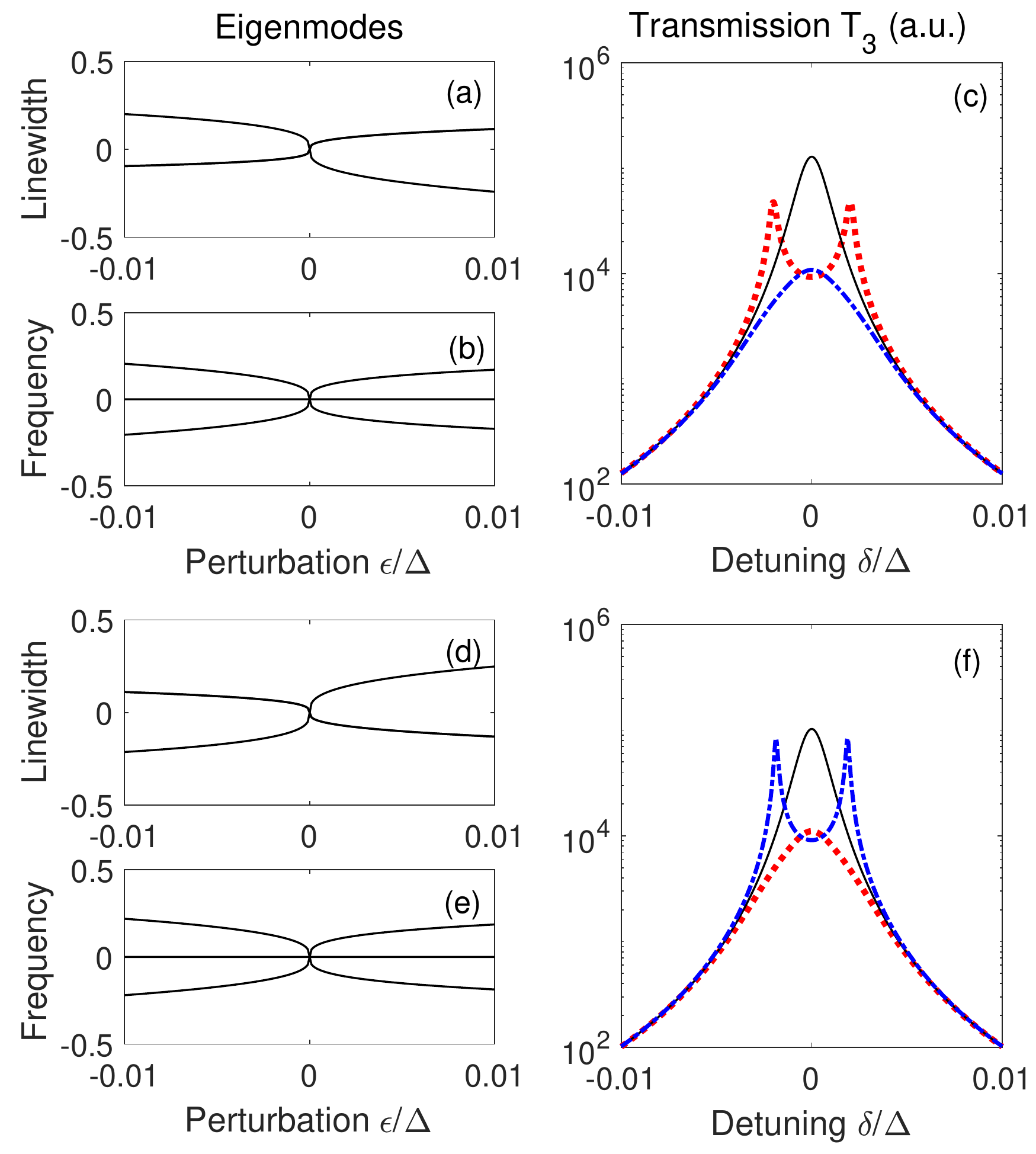} \\	
\caption{\small The eigenmodes (left) and the transmission spectra (right) near the twin critical-coupling EP$_3$ [corresponding to (a)-(c) and (d)-(f), respectively] designed at $\sigma=0$. The black solid curves in the transmission spectra are those at the EP, while the red dotted (blue dash-dotted) curves are the spectra with perturbation $\epsilon/\Delta=-10^{-8}$ ($\epsilon/\Delta=10^{-8}$) on $\gamma_3/\Delta$. The loss (gain) rates of the resonators are chosen as $\gamma_1/\Delta=-1.3093$, $\gamma_2/\Delta=-0.6547$ and $\gamma_3/\Delta=1.9640$.}\label{fig:EP3c}
\end{figure}

As shown in Fig.~\ref{fig:EP3c}, the mode frequencies are split into three different values on both sides of the EP$_3$ at the origin, while two of the three linewidths merge together. Take the first twin EP$_3$ for example, when $\epsilon<0$, the two modes with nonzero and opposite detunings have identical linewidth [see the lower curve left to the origin in Fig.~\ref{fig:EP3c}(a)]. The linewidth of the third mode with zero frequency [see the upper curve left to the origin in Fig.~\ref{fig:EP3c}(a)] is broader and hence is almost unobsevable in the transmission spectrum. Thus, we see only two narrow-linewidth peaks [see the red dotted curve in Fig.~\ref{fig:EP3c}(c)]. On the contrary, when $\epsilon>0$, only the mode with zero frequency has narrow linewidth that produces the peak in the transmission spectrum [see the blue dash-dotted curve in Fig.~\ref{fig:EP3c}(c)], while the other two detuned modes having identical broader linewidths do not produce observable peaks.

Comparing with the previously studied EP$_2$, the transmission property near EP$_3$ is much more sensitive to the perturbation $\epsilon$. In the simulated transmission spectra, we choose $\epsilon=10^{-8}\Delta$, which is three orders smaller than that in the EP$_2$ simulations, and still observe sharp splitted peaks. The mode splitting is expectation to be proportional to $\epsilon^{1/3}$ instead of $\epsilon^{1/2}$ near EP$_2$, which can be verifed fro the observed value $10^{-3}\sim10^{-2}\Delta$ from the spectra. This extraordinary sensitivity has been experimentally approved for ultra-sensitive optical detection with three resonators \cite{Chen2017}.

\begin{figure}
\begin{center}
\includegraphics[width=0.8\columnwidth]{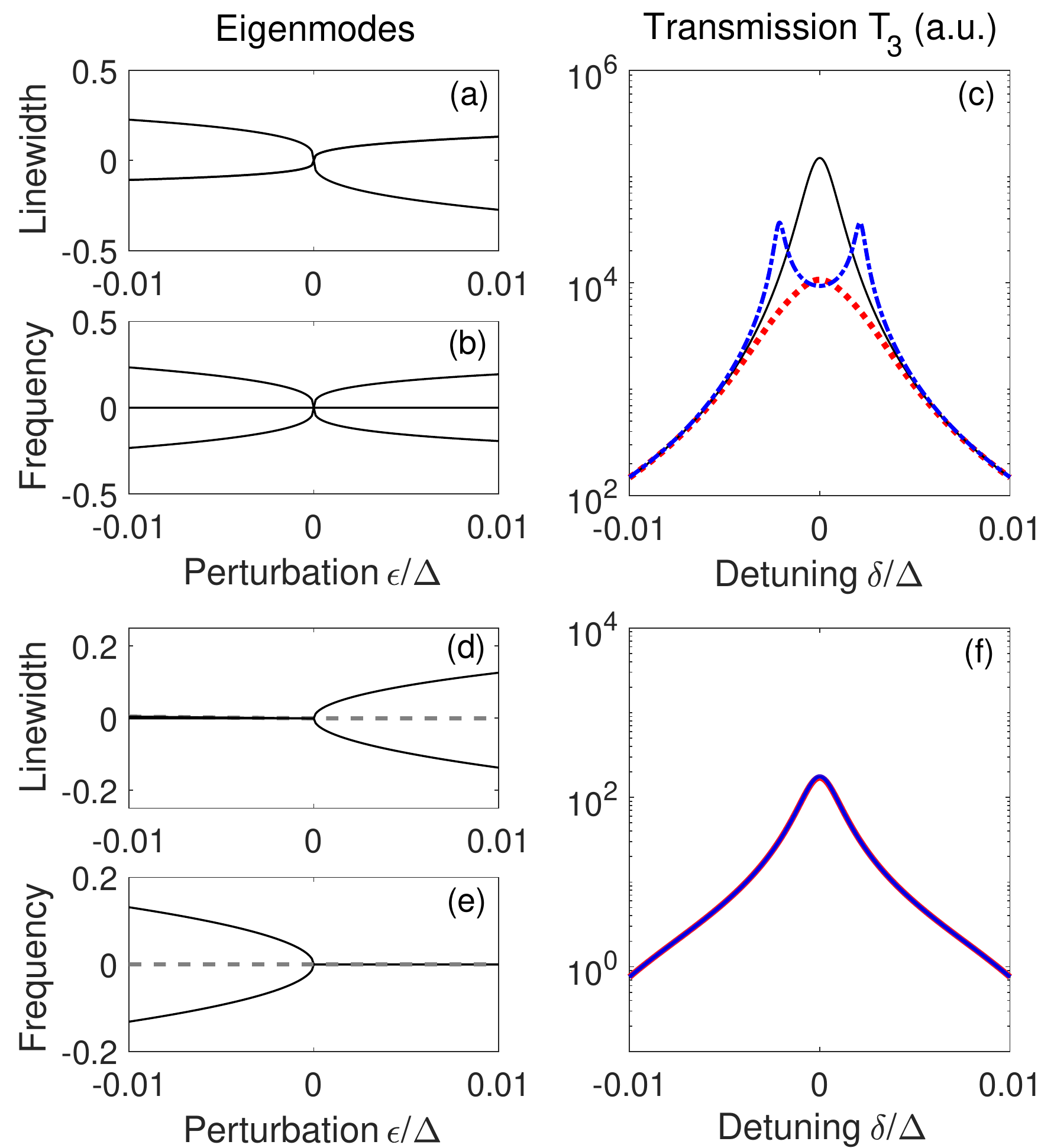}
\end{center}	
\caption{\small The eigenmodes (left) and the transmission spectra (right) near the twin EPs in critical coupling regime, in which one is an EP$_3$ [(a)-(c)] and the other is an EP$_2$ [(d)-(f)]. The black solid curves in the transmission spectra are those at the EP, while the red dotted (blue dash-dotted) curves are the spectra with perturbation $\epsilon/\Delta=-10^{-8}$ ($\epsilon/\Delta=10^{-8}$) on $\gamma_3$. The loss (gain) rates of the resonators are chosen as $\gamma_1/\Delta=-1.7321$, $\gamma_2/\Delta=0$ and $\gamma_3/\Delta=1.7321$. }\label{fig:DP3c}
\end{figure}

To illustrate the possible transition from EP$_3$ to EP$_2$ indicated in Section \ref{Sec:Model-EP3}, we also simulate  the networks with $\gamma_1/\Delta=-1.7321$, $\gamma_2/\Delta=0$ and $\gamma_3/\Delta=1.7321$, whose coupling constants are also listed in Table \ref{Tab3}. Consistent with our prediction, the second EP is second-order and is realized by an actual two-resonator network realization with the second resonator being isolated (i.e.,  $\kappa_{12}=\kappa_{23}=0$). As shown by Fig.~\ref{fig:DP3c} (lower plots), the variance of the eigenmodes is very similar to that of EP$_2$'s in Fig.~\ref{fig:EP2c}. In the transmission spectrum Fig.~\ref{fig:EP3c}(f), mode splitting is not discernable under the same perturbation $\epsilon/\Delta=10^{-8}$, which  verifies the resulting EP$_2$ does not exhibit the same high sensitivity as that with the EP$_3$.

\section{CONCLUSIONS}\label{Sec:five}
In this paper, we studied the engineering of exceptional points in three-resonator networks, which many interesting structures. A full classification is provided, which is according to the average coupling strength, for all physically admissible EPs. All the required formulas for calculating the required coupling constants are also provided, with which one can design the network with an arbitrary given set of three resonators and any desired EP that is admissible.
			
In the literature, experiments with EP$_2$ and EP$_3$ in critical and weak coupling regimes have been reported, which are all special cases of this paper. However, we did not see any study on the EP$_2$ in strong coupling regime associated with circularly coupled resonator networks. Although the presence of backscattering is sometimes annoying in practice, the physics behind may be interesting because all associated eigenmodes are lossless, which exhibits a completely different energy balance mechanism with that of EP$_2$'s
in the weak coupling regime where the energy is balanced between losses and gains. We expect that such difference may lead to new applications.

The chirality of EP's in three-resonator networks is another interesting topic that can be explored in the future. In our  simulations, some symmetry relations are observed between the twin EPs associated with each EP eigenvalue, e.g., between Figs.~\ref{fig:EP2c}(a-b) and Figs.~\ref{fig:EP2c}(d-e) in the weak-coupling regime. We conjecture that this is related with the chirality of their associated eigenmodes, which is still not clear to the authors. Also, in order to avoid unwanted backscattering under many circumstances, it will be interesting to study how to further manipulate the chiral optical modes travelling in the circularly coupled networks that host EP's in the strong-coupling regime.
			
The full classification of EP's in multi-resonator networks are much more complicated. The resulting characteristic polynomials are much more complicated if without any symmetry assumptions. This will be an important direction to be explored in the future.


\appendix
\section{Analysis of EP$_2$ in weak coupling regime}\label{app:Bp}
To analyze the EPs, we start from the characteristic polynomial for the matrix $A$ in the balanced loss-gain case:
$$P(s) ={\rm det}(s\mathbb{I}_3-A)= s^3 + a_1 s +a_0=0,$$
which can be derived from Eq.~(\ref{}) for $n=3$, where
\begin{eqnarray}\label{eq:Delta3}
a_0	&=& \gamma_1 \gamma_2 \gamma_3 + \gamma_3 \kappa_{12}^2 + \gamma_1 \kappa_{23}^2 + \gamma_2 \kappa_{31}^2 \nonumber \\
&& +
2 \imath \kappa_{12} \kappa_{23} \kappa_{31}, \\
a_1	&=& \gamma_1 \gamma_2 + \gamma_1 \gamma_3 + \gamma_2 \gamma_3 \nonumber\\
&& + \kappa_{12}^2 + \kappa_{23}^2 + \kappa_{31}^2.
\end{eqnarray}

First, it is easy to prove that, in the balanced loss-gain case (i.e., $\gamma_1+\gamma_2+\gamma_3=0$),
\begin{equation}
\gamma_1\gamma_2+\gamma_2\gamma_3+\gamma_3\gamma_1=-3\Delta^2.
\end{equation}
On the other hand, when the system has an EP at $\sigma$, the characteristic polynomial can also be written as
$$P(s)=(s-\sigma)^2(s+2\sigma)=s^3-3\sigma^2s+\sigma^3.$$
Therfore, by comparing it with $\Delta(s)$, we have
\begin{equation}\label{eq:a3}
\begin{split}
2\sigma^3 &= \gamma_1 \gamma_2 \gamma_3 + \gamma_1 \kappa_{23}^2 + \gamma_2 \kappa_{31}^2 + \gamma_3 \kappa_{12}^2 ,\\
-3\sigma^2 &=  \gamma_1 \gamma_2 + \gamma_1 \gamma_3 + \gamma_2 \gamma_3 \,
 + \kappa_{23}^2 + \kappa_{31}^2,
\end{split}
\end{equation}
where the imaginary part of $a_0$ is $\kappa_{12}\kappa_{23}\kappa_{31}=0$. This means that at least one of the coupling constants is zero, i.e., EPs in the weak-coupling regime exist only in linearly coupled networks. For example, when $\kappa_{12}=0$, Eq.~(\ref{eq:a3}) can be taken as linear equations of $\kappa_{23}^2$ and $\kappa_{31}^2$, from which we obtain the following parametrization of coupling constants:
\begin{eqnarray}
\kappa_{13} &=&  (\gamma_1+\sigma)\sqrt{\frac{\gamma_1-2\sigma}{\gamma_1-\gamma_2}},\label{eq:EP2c1}\\
\kappa_{23} &=& (\gamma_2+\sigma)\sqrt{\frac{\gamma_2-2\sigma}{\gamma_2-\gamma_1}}.\label{eq:EP2c2}
\end{eqnarray}

To guarantee that both $\kappa_{13}$ and $\kappa_{23}$ are real numbers, it is easy to derive from the above expressions that the EP eigenvalue $\sigma$ is bounded by
\begin{equation}\label{eq:EP2 range}
\frac{\gamma_1}{2}<\sigma < \frac{\gamma_2}{2}.
\end{equation}
Similarly, one can compute the coupling constants for the other two network topologies with $\kappa_{23}=0$ and $\kappa_{31}=0$, whose corresponding EP eigenvalues satisfy $\frac{\gamma_2}{2}<\sigma < \frac{\gamma_3}{2}$ and $\frac{\gamma_1}{2}<\sigma < \frac{\gamma_3}{2}$, respectively. These EPs correspond to the three horizontal line segments shown in Fig.~\ref{fig:EPdistribution}.

Note that the above analysis only shows that the system's eigenvalues are degenerate at $\sigma$. Whether it also leads to the degeneracy of eigenvectors needs to be verified by the analysis of the Jordan form that can be easily done. Our calculation shows that for all the above physically admissible degenerate eigenvalues, the resulting $A$ is always associated with an EP but not DP.

\section{Analysis of EP$_2$ in strong coupling regime}\label{app:A}
Similar to the coefficient analysis of $\Delta(s)$ in the weak coupling regime, we have
\begin{equation}\label{eq:EP2r}
\begin{split}
3\sigma^2 &=  \gamma_1 \gamma_2 + \gamma_1 \gamma_3 + \gamma_2 \gamma_3 \,
+ \kappa_{12}^2 + \kappa_{23}^2 + \kappa_{31}^2, \\
0&=  \gamma_1 \gamma_2 \gamma_3+\gamma_3 \kappa_{12}^2 + \gamma_1 \kappa_{23}^2 + \gamma_2 \kappa_{31}^2,\\
\sigma^6 &=  \kappa_{12}^2 \kappa_{23}^2 \kappa_{31}^2.
\end{split}
\end{equation}
The last equality shows that all the coupling constants are nonzero, i.e., EP exist only in circularly coupled networks in which the three resonator all couple to each other. This is a major topological difference with EPs in the weak coupling regime.

To calculate the coupling constants associated with a potential EP, we have to solve the above nonlinear equations in terms of $\kappa_{12}^2$, $\kappa_{12}^2$ and $\kappa_{12}^2$. This is much more complicated than the case of weak-coupling EP's, and will be discussed in the following two sub-classes.

\subsection{Symmetric network}
When two resonators have identical loss rates, say $\gamma_1=\gamma_2=\gamma\neq 0$ and $\gamma_3=-2\gamma$, we have
\begin{equation}\label{eq:EP2rsym}
\begin{split}
3\sigma^2+3 \gamma^2 &= \kappa_{12}^2 + \kappa_{23}^2 + \kappa_{31}^2, \\
2\gamma^2&=  -2 \kappa_{12}^2 + \kappa_{23}^2 + \kappa_{31}^2,\\
\sigma^6 &=  \kappa_{12}^2 \kappa_{23}^2 \kappa_{31}^2,
\end{split}
\end{equation}
from which one pcan easily solve
\begin{eqnarray}
\kappa_{12} &=& \sqrt{\frac{\gamma ^2}{3}+\sigma ^2}, \\
\kappa_{23} &=& \sqrt{\frac{4 \gamma ^2}{3}+\sigma ^2\pm\frac{\gamma( 4 \gamma^2+9 \sigma ^2)}{3 \sqrt{\gamma ^2+3 \sigma ^2}}}, \\
\kappa_{31} &=& \sqrt{\frac{4 \gamma ^2}{3}+\sigma ^2\mp\frac{\gamma(4 \gamma^2+9  \sigma ^2)}{3\sqrt{ \gamma ^2+3 \sigma ^2}}}.
\end{eqnarray}
Because the first and the second resonators have identical loss rates and frequencies, these two solutions actually correspond to the same circularly coupled networks.

\subsection{Asymmetric network}
In asymmetric networks, the loss rates are mutually different from each other. In such case, there are generally no analytical solutions to these equations.
We firstly solve from the first two equations of (\ref{eq:EP2r}) that
\begin{eqnarray}
\kappa_{12}^2 &=& \frac{\gamma_1-\gamma_2}{\gamma_3-\gamma_1}\left[{\kappa_{31}^2}  -\frac{\gamma_1 (\gamma_1^2 + 3 \sigma^2)}{\gamma_1-\gamma_2}\right],\\
\kappa_{23}^2 &=& \frac{\gamma_2-\gamma_3}{\gamma_3-\gamma_1}\left[{\kappa_{31}^2}  -\frac{\gamma_2 (\gamma_2^2 + 3 \sigma^2)}{\gamma_2-\gamma_3}\right],
\end{eqnarray}
which can be replaced into the third equation to obtain the following cubic equation for $\kappa_{12}^2$:
\begin{equation}\label{eq:cubic}
f(x)=  x(x  - x_1)(x-x_2)=y_0,
\end{equation}
where, under the assumption $\gamma_1<\gamma_2<\gamma_3$,
\begin{eqnarray*}\label{}
x_{1}  &=& \frac{\gamma_3 (\gamma_3^2 + 3 \sigma^2)}{\gamma_3-\gamma_2}>0,\\
x_{2}  &=& \frac{\gamma_1 (\gamma_1^2 + 3 \sigma^2)}{\gamma_1-\gamma_2}>0,\\
y_0  &=&\frac{(\gamma_3 - \gamma_1)^2\sigma^6}{(\gamma_1-\gamma_2)(\gamma_2-\gamma_3)}>0.
\end{eqnarray*}

\begin{figure} \label{fig:EP2r_exist}
\center
{\includegraphics[width=0.6\columnwidth]{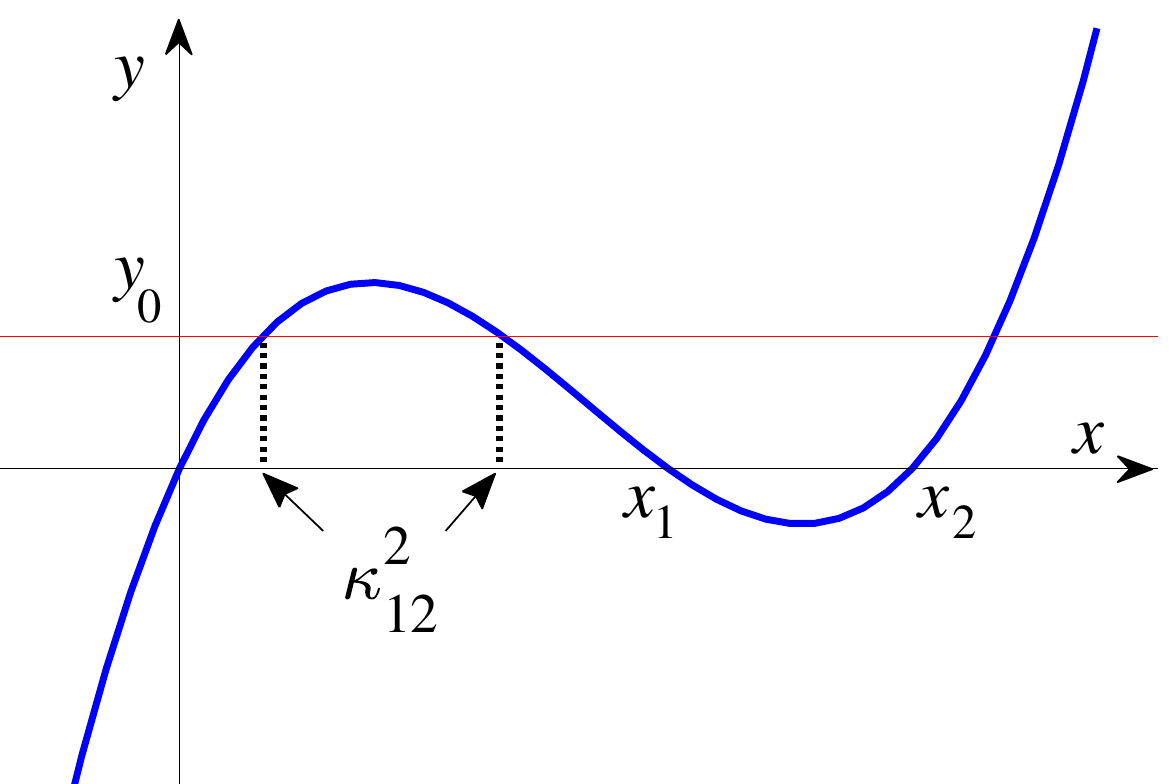}
	}	
\caption{The existence of EP$_2$ in circular networks, where the cross of the cubic curve and the horizontal line implies that Eq.~(\ref{eq:cubic}) always has two positive roots smaller than both $x_1$ and $x_2$. They correspond to $\kappa_{31}^2$ associated with the two EP realizations.}
\end{figure}

We can see that a physically admissible EP$_2$ exists, for the positivity of $\kappa_{12}^2$, $\kappa_{23}^2$ and $\kappa_{31}^2$, if and only if the cubic equation has a  solution $x>0$ that is smaller than both $x_1$ and $x_2$, like what is shown in Fig.~\ref{fig:EP2r_exist}. The figure also shows that there must be two distinct solutions if the solution exists,

The existence condition requires that the three roots of the cubic equation must all be real. Rewrite (\ref{eq:cubic}) as $f(x)=x^3+bx^2+cx+d=0$, the condition is equivalent to that the discriminant
$$\Delta = 18  b c d - 4 b^3 d + b^2 c^2 - 4  c^3 - 27 d^2>0,$$
which, after a tedious calculation, can be shown to be the case, as follows
\begin{eqnarray*}
  \Delta &=& (\gamma_3-\gamma_1)^2 \left[\left(\frac{\Delta^2\sigma}{3}\right)^2 +\gamma_1^2 \gamma_2^2 \gamma_3^2\right]^2\\
   && \cdot\left[\left(\frac{2\Delta^2\sigma}{3}\right)^2+12 \Delta^2\sigma ^4+108 \sigma ^6+\gamma_1^2 \gamma_2^2 \gamma_3^2\right]>0.
\end{eqnarray*}

Therefore, we conclude that for any given set of resonators and any desired degenerate eigenvalues $\imath\sigma$, there are always two distinct network realizations. We are not able to analytically verify that they are all true EP$_2$'s, but a large amount of numerical tests show that they are all  associated with EP's instead of DP's.

\acknowledgments
This work is supported by NSFC grants (Nos. 61773232, 61374091 and 61134008) and National Key
Research and Development Program of China (Grant
No. 2017YFA0304300).


\end{document}